\documentclass[aps,twocolumn,showpacs,preprintnumbers,prb,amsmath,amssymb,amsfonts,superscriptaddress,floatfix]{revtex4-1}
\usepackage{graphicx}
\usepackage{dcolumn}
\usepackage{bm}
\usepackage[version=3]{mhchem} 
\usepackage{color}
\usepackage{array}
\usepackage{multirow}

\begin{document}
\title{Applicability of the strongly constrained and appropriately normed
density functional to transition metal magnetism}

\author{Yuhao Fu}

\author{David J. Singh}
\email{singhdj@missouri.edu}

\affiliation{Department of Physics and Astronomy, University of Missouri, Columbia, MO 65211-7010 USA}
\date{\today}

\begin{abstract}
We find that the recently developed
self consistent and appropriately normed
(SCAN) meta-generalized gradient approximation,
which has been found to provide highly accurate results
for many materials, is, however, not able to
describe the stability and
properties of phases of Fe important for steel. This is
due to an overestimated tendency towards magnetism and exaggeration
of magnetic energies, which we also
find in other transition metals.
\end{abstract}

\maketitle

Density functional theory (DFT) calculations \cite{kohn}
are a central tool in condensed matter physics, chemistry, and
materials science. This utility is the result of the availability of
sufficient accuracy in tractable approximate functionals. This
enables predictive calculations of properties of interest and 
elucidations of underlying mechanisms of physical behavior.
Therefore the development of new practical
functionals that improve the accuracy,
and therefore the range of behaviors and materials that can be
studied with DFT calculations, is of great interest.

Steel is arguably the most important industrial material.
Annual production exceeds 1.7 billion metric tonnes.
Steels are complex materials whose properties are controlled by
microstructure. These microstructures are what provides steel
with desirable combinations of ductility, toughness and tensile strength.
These microstructures come from balances between
different phases mainly in the Fe-C phase diagram. \cite{bramfitt}
While the ground state of Fe is body centered cubic (bcc),
an an equilibrium face centered cubic (fcc) phase
exists between between 1185 K and 1667 K. Carbon has a much higher solubility
in this fcc phase (up to 2.14 wt\% and 0.76 wt\% at the eutectoid)
than in the bcc phase (maximum of
0.022 wt\%), leading to an easily accessed
eutectoid point in the phase diagram (at
1000 K and 0.77 wt\% C). Cooling leads
to nanoscale and microscale precipitation
of cementite (Fe$_3$C, a very hard phase), in a bcc Fe matrix,
as well as non-equilibrium austenite (fcc Fe with C)
and sometimes other phases associated
with alloying elements, to form microstructures such as perlite,
martensite and bainite. These microstructures, sometimes modified by
mechanical deformation steps, are key to the properties of steel.
First principles based understanding of steel requires
the ability to model these different phases and their relationships, most
importantly the relationship between the ground state bcc structure (ferrite)
and the fcc structure (austenite).

This has posed ongoing challenges to density functional calculations.
Early on it was found that the otherwise highly successful local
(spin) density approximation (LDA), cannot describe Fe. In particular,
it was shown that the LDA predicts a non-magnetic fcc ground state for
Fe, with the ferromagnetic bcc structure lying higher in energy.
\cite{wang}
The LDA does, however, provide an accurate value of the spin magnetization
of Fe, when constrained to its experimental bcc structure.

An important step was the development of generalized gradient
approximation (GGA) functionals,
\cite{lm-gga,pw86,pw91,pbe}
based on knowledge of the behavior of the
exchange correlation hole
in inhomogeneous electron gasses. \cite{adiabatic,jones}
In addition to correctly predicting the bcc ground state and
spin magnetization of Fe,
\cite{pw91,singh-fe,amador-fe,zhu-fe,asada-fe,haglund}
these GGA functionals greatly improved the
energetics of a wide variety of molecules and solids. This was a remarkable
achievement, especially considering that these
GGA functionals were based on constraints and scaling for
the electron gas and not fits to known materials properties.

Therefore, it is very reasonable to assume that functionals that
incorporate additional known exact properties of the inhomogeneous
electron gas will at least on average improve the description of
atoms, molecules and solids.
A significant recent development along these
lines was the construction of a
strongly constrained and appropriately normed (SCAN) functional. \cite{sun}
This is a semi-local meta-GGA
functional. Meta-GGA functionals are more convenient
for calculations than hybrid functionals, \cite{b3lyp,HSE02}
especially in extended systems.

The SCAN functional satisfies exact constraints, including importantly
the Lieb-Oxford lower bound for the exchange energy,
\cite{const-perdew,const-lieb}
also important for
the construction of the earlier GGA functionals, as well as scaling
relations.
\cite{const-levy}
It is also designed to revert to the LDA for the uniform electron
gas (a norm) and also uses the hydrogen atom as a norm for the
exchange. This is important in regards to self-interaction errors.
It is designed to be accurate both for the slowly varying electron gas,
and for atoms, which is not possible in GGA functionals. \cite{const-p2}

Tests done to date generally confirm the expectation that SCAN provides
highly accurate results for many materials,
\cite{sun,sun2,tran,zhang,isaacs}
as might be expected from the many constraints that it satisfies.
\cite{singh-harbola}
However, there is at least one
indication
that SCAN may not improve the already
generally good description of magnetism in some metallic ferromagnets.
Isaacs and co-workers \cite{isaacs}
reported that the magnetization of
Fe, Co and Ni are enhanced by 0.42 $\mu_B$, 0.13 $\mu_B$ and 0.1 $\mu_B$,
respectively,
relative to the widely used GGA functional of
Perdew, Burke and Ernzerhof (PBE). \cite{pbe}
They observed that this degrades
agreement with experiment for Fe and Ni.
Ekholm and co-workers, also performed calculations for Fe, Co and Ni, and
found that the moments were enhanced relative to experiment, which they
ascribed to a downshift of the 3d states.
\cite{ekholm}

We did calculations with the LDA, the PBE GGA and the SCAN functional
using two different methods, specifically the projector augmented wave (PAW)
method \cite{PAW} as implemented in the VASP code, \cite{VASP}
and the all electron general
potential linearized augmented planewave (LAPW) method, \cite{singh-lapw}
as implemented in the
WIEN2k code. \cite{wien2k}
The VASP code includes a self-consistent calculation with the SCAN functional,
except that that it relies on PAW potentials constructed for the
PBE GGA, which is an approximation.
The LAPW method as implemented in WIEN2k is an all electron method that
does not rely on pseudopotentials. However, at present, SCAN calculations
with this method must be done non-self-consistently, in particular,
calculating the energy using the SCAN functional, but based on the 
density from a semi-local calculation. We used the
PBE GGA with the constrained DFT,
\cite{dederichs}
specifically the fixed spin moment (FSM)
procedure,
\cite{fsm1,fsm2,moruzzi}
to generate the spin densities for calculating the SCAN total
energies.

This procedure involves solving the Kohn-Sham equations with a constraint
that the integrated spin density (the spin-moment) equal a specified
value. This is achieved by
imposing the constraint via a difference in spin-up and spin-down
Fermi levels, equivalent to a magnetic field operating on spin only.
\cite{wagner}
We used dense grids of discrete moments to obtain the plots shown here.
This allows us also to calculate the total energy as a function
of the constrained moment for ferromagnetic materials, and provides
insights into the problems in the treatment
of magnetic transition metals with SCAN.
We carefully converged the calculations, using large basis sets, and dense
convergence tested k-point grids for all materials.
We compared the results from the two codes and
find very similar results, which supports the different approximations involved.
We also did
self consistent calculations including spin orbit for the PBE and LDA
functionals to quantify the effect of spin orbit, which could not be
applied in FSM calculations for the SCAN functional.
These show that the effects of spin orbit are small on the scale of the
differences between the functionals, and cannot resolve the discrepancies.

\begin{figure}[tbp]
\includegraphics[width=0.94\columnwidth,angle=0]{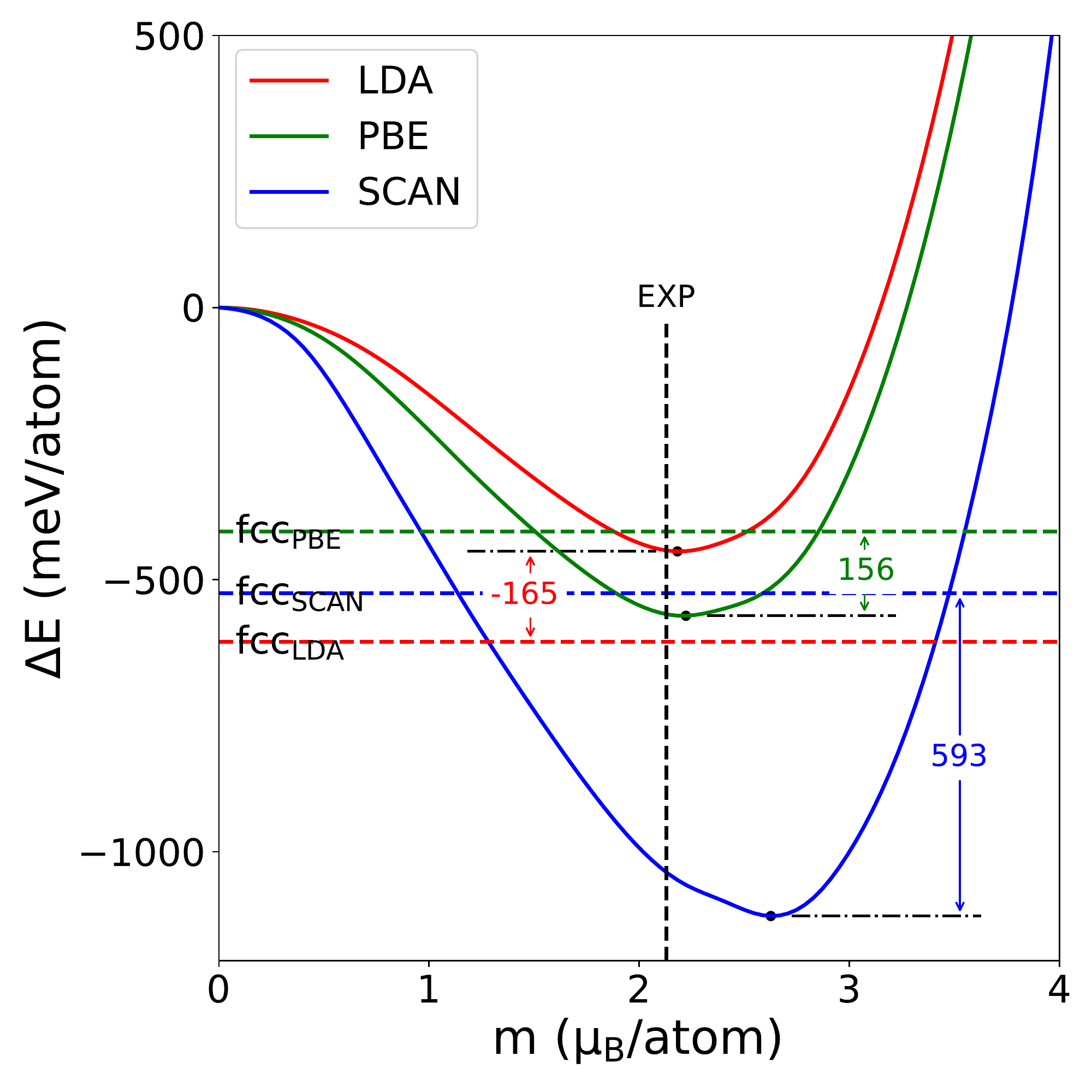}
\caption{FSM energy for bcc Fe at the experimental lattice
constant of 2.86 \AA, on a per atom basis. The dashed lines are the energies
of non-spin-polarized fcc Fe, at the optimized lattice parameter
for the different functionals. The small dots indicate the minimum energy
points.}
\label{fsm-fe-bcc}
\end{figure}

The measured saturation magnetizations of
Fe, Co and Ni are 2.22 $\mu_B$, 1.72 $\mu_B$ and 0.62 $\mu_B$,
on a per atom basis.
\cite{danan,myers}
These include both spin and orbital contributions.
The orbital moments of Fe and Co from
x-ray magnetic circular dichroism (XMCD) experiments are
0.09 $\mu_B$ and 0.15 $\mu_B$, per atom,
\cite{chen-xmcd}
while the experimental value for Ni is 0.05 $\mu_B$.
\cite{lb}
Our spin ($m_{sp}$) and orbital
($m_{orb}$) moments, at the experimental
lattice parameters from LAPW
calculations with the PBE functional including
spin orbit are
$m_{sp}$=2.22 $\mu_B$ and $m_{orb}$=0.04 $\mu_B$ for Fe,
$m_{sp}$=1.62 $\mu_B$ and $m_{orb}$=0.08 $\mu_B$ for Co, and
$m_{sp}$=0.63 $\mu_B$ and $m_{orb}$=0.05 $\mu_B$ for Ni,
i.e. spin moments very close to the experimental values, and orbital
moments are small and underestimated for Fe and Co,
as in prior calculations. \cite{daalderop}
In our calculations without spin orbit, the PBE spin moments
are $m_{sp}$=2.22 $\mu_B$, $m_{sp}$=1.62 $\mu_B$, and $m_{sp}$=0.63 $\mu_B$,
for Fe, Co and Ni, respectively, which are the same as those with
spin orbit to the quoted precision. Thus spin orbit does not have
a significant effect on the calculated spin moments for these 3d ferromagnets.
It is also to be noted that any enhancement of the spin moment over
the PBE values will degrade agreement with experiment, including the case
of Co.

Fig. \ref{fsm-fe-bcc} shows our results for the magnetic energy of bcc Fe
at its experimental lattice parameter, in comparison with the
energy of non-spin-polarized fcc Fe. Numerical values and
magnetic moments are given in Table \ref{tab-fe}.
As seen, the SCAN functional
yields dramatically different results from the LDA and PBE functionals.
Energy plays a central role in density functional theory.
As mentioned, the LDA fails for Fe, predicting that the fcc structure
has lower energy, in particular by 0.165 eV.
The PBE functional yields the correct ordering, with an energy difference of
0.156 eV, considering a non-magnetic fcc structure.
The SCAN functional predicts a much more stable bcc structure,
with an overestimated spin moment of 2.63 $\mu_B$/atom and an fcc -
bcc energy difference of 0.593 eV. This is due to a much
larger magnetic energy.
Self consistent calculations using VASP yield similar numbers,
specifically a spin moment of 2.65 $\mu_B$ and an energy difference
of 0.579 eV, for SCAN.
While these numbers do not include the magnetic
enthalpy of fcc Fe, it is clear that SCAN predicts an overly stable
ferromagnetic state for bcc Fe.
The experimental enthalpy difference between bcc and fcc Fe at 1185 K from
assessed calorimetric measurements is 0.009 eV/atom,
while the low temperature energy difference from thermodynamic
models based on experimental data is 0.06 eV/atom. \cite{chen}

\begin{table}
\caption{Calculated properties of Fe. $a_{exp}$ and $a_{calc}$
are the experimental and calculated
lattice parameters of bcc Fe, respectively.
The fcc-bcc energy difference $\Delta$$E_{fcc-bcc}$ is as in
Fig. \ref{fsm-fe-bcc}. $\Delta$$E_{mag}$ is the magnetic energy from
the difference between non-spin polarized and ferromagnetic
states. Energies are per atom.}
\begin{tabular}{lcccc}
\hline
   & ~~~LDA~~~ & ~~~PBE~~~ & ~~SCAN~~ & ~~Expt.~~ \\
\hline
$a$ (\AA)                    & 2.76   & 2.84   & 2.85   & 2.86     \\
$m_{sp}$($a_{exp}$) ($\mu_B$)  & 2.21   & 2.21   & 2.63   & 2.13   \\
$m_{sp}$($a_{calc}$) ($\mu_B$) & 2.00   & 2.16   & 2.60   &  -    \\
$m_{orb}$($a_{exp}$) ($\mu_B$) & 0.05   & 0.04   &  -     & 0.09  \\
$\Delta$$E_{mag}$($a_{exp}$) (meV)  & 448 & 566 & 1117 & - \\
$\Delta$$E_{mag}$($a_{calc}$) (meV) & 317 & 529 & 1078 & - \\
$\Delta$$E_{fcc-bcc}$ (meV) & -165 & 156 & 593 & 60$^a$ \\
\hline
\end{tabular}
$^a$estimate from extrapolated thermodynamic data (see text).
\label{tab-fe}
\end{table}

\begin{figure}[tbp]
\includegraphics[width=0.90\columnwidth,angle=0]{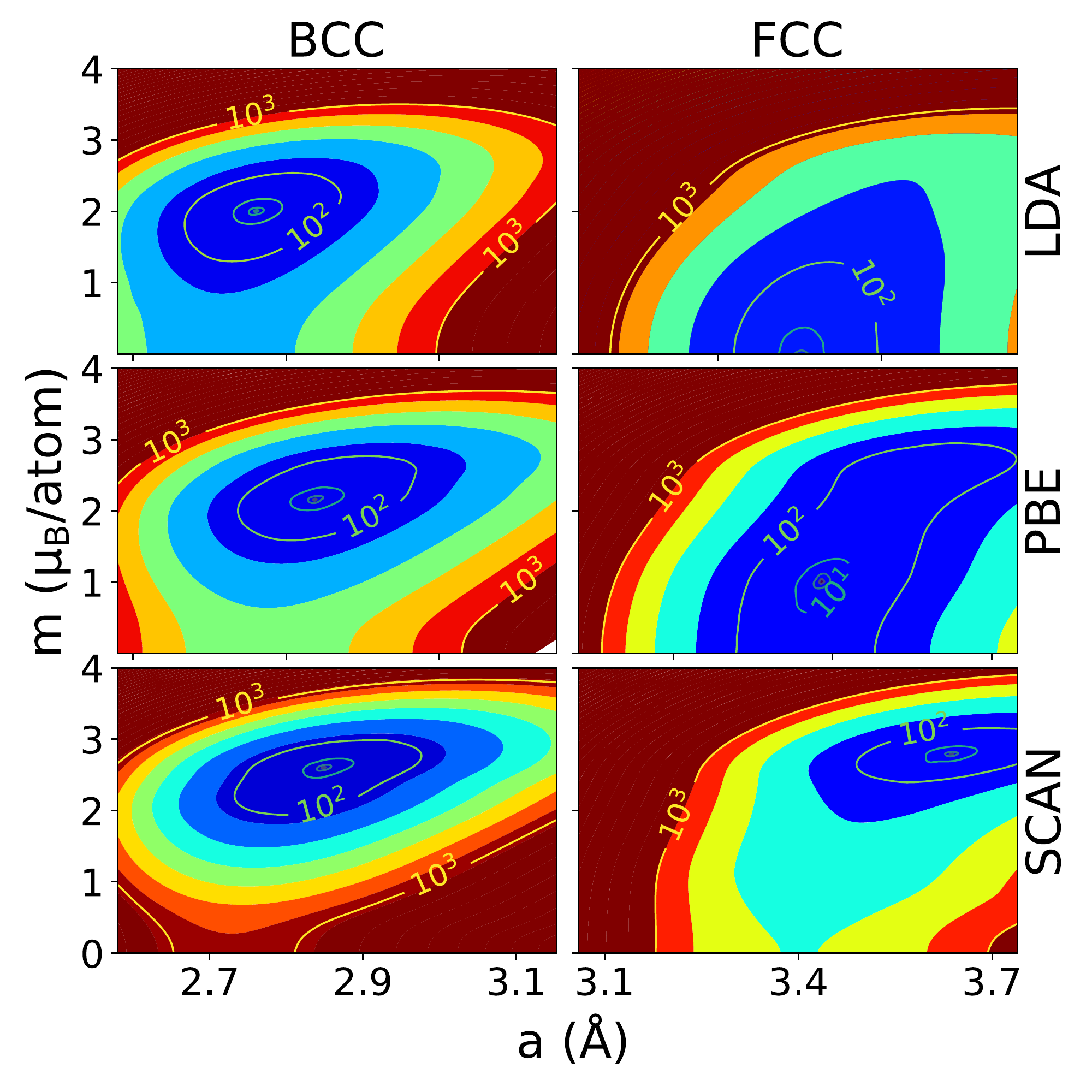}
\caption{LDA, PBE and SCAN
FSM energy in meV/atom for bcc and fcc Fe as functions of lattice
parameter and spin moment.}
\label{fe-map}
\end{figure}

Fig. \ref{fe-map} shows the FSM energy as functions of lattice
parameter and moment for bcc and fcc Fe.
In accord with older work, \cite{wang,haglund}
in addition to its failure to predict the correct ground state,
the LDA strongly underestimates the lattice parameter of magnetic
bcc Fe, while the PBE GGA give values in closer agreement with experimental
data. The SCAN functional gives a lattice parameter similar to PBE
for the bcc structure.
The SCAN functional predicts very different behavior for the fcc phase.
When constrained to ferromagnetism, the LDA and PBE
predict either no magnetism or a low moment state. The SCAN functional
predicts a high moment state. While high moment ferromagnetism does not
preclude a still lower energy ground state with antiferromagnetism, it
is incompatible with a weak low moment antiferromagnetic state, due to the
large magnetic energy associated with the high moment state.

Experimental information on the magnetism of free fcc Fe is limited by the
fact that it is not a stable low temperature phase.
However, fcc Fe films grown epitaxially on Cu are paramagnetic at
ambient temperature, and become antiferromagnetic at low
temperature with $T_N$$\sim$65 K,
\cite{macedo}
similar to the behavior of small fcc Fe precipitates in an fcc Cu matrix.
\cite{gonser}
According to neutron diffraction measurements these have a small
moment of $\sim$0.5$\mu_B$ per Fe.
\cite{tsunoda}
Based on this, on this, as well as the properties of
non-ferromagnetic austenitic steels, \cite{gonser}
thermodynamic modeling,
and extrapolation of alloy data \cite{chen,acet} it is thought
that fcc Fe is an itinerant weak
antiferromagnet with a Neel temperature below 70 K, and a
relatively small contribution of magnetism to the energy.
DFT studies have indicted that there is an additional high volume 
high spin ferromagnetic state with
higher energy, and this has been discussed in connection
with the stability of the fcc phase between 1185 K and 1667 K.
\cite{haglund,chen}

We also did self consistent calculations with VASP for the energy
and moments of a hypothetical antiferromagnetic bcc Fe,
where the moments of the two Fe atoms in the conventional cubic cell
are oppositely aligned. We find that with the PBE
functional the moments as measured by the
spin density around Fe sites, is reduced from 2.25 $\mu_B$ in ferromagnetic
case (note there is a small negative interstitial spin moment of $\sim$  -0.03
$\mu_B$) to 1.71 $\mu_B$. In contrast, the SCAN result for
the antiferromagnetic case of 2.66 $\mu_B$
is almost exactly the same as for the ferromagnetic
case, i.e. 2.65 $\mu_B$. PBE predicts intermediate itinerant / local moment
behavior for bcc Fe, while SCAN predicts that Fe is in the local moment
limit, in general disagreement with experiment.
\cite{heine}

\begin{figure}[tbp]
\includegraphics[width=0.95\columnwidth,angle=0]{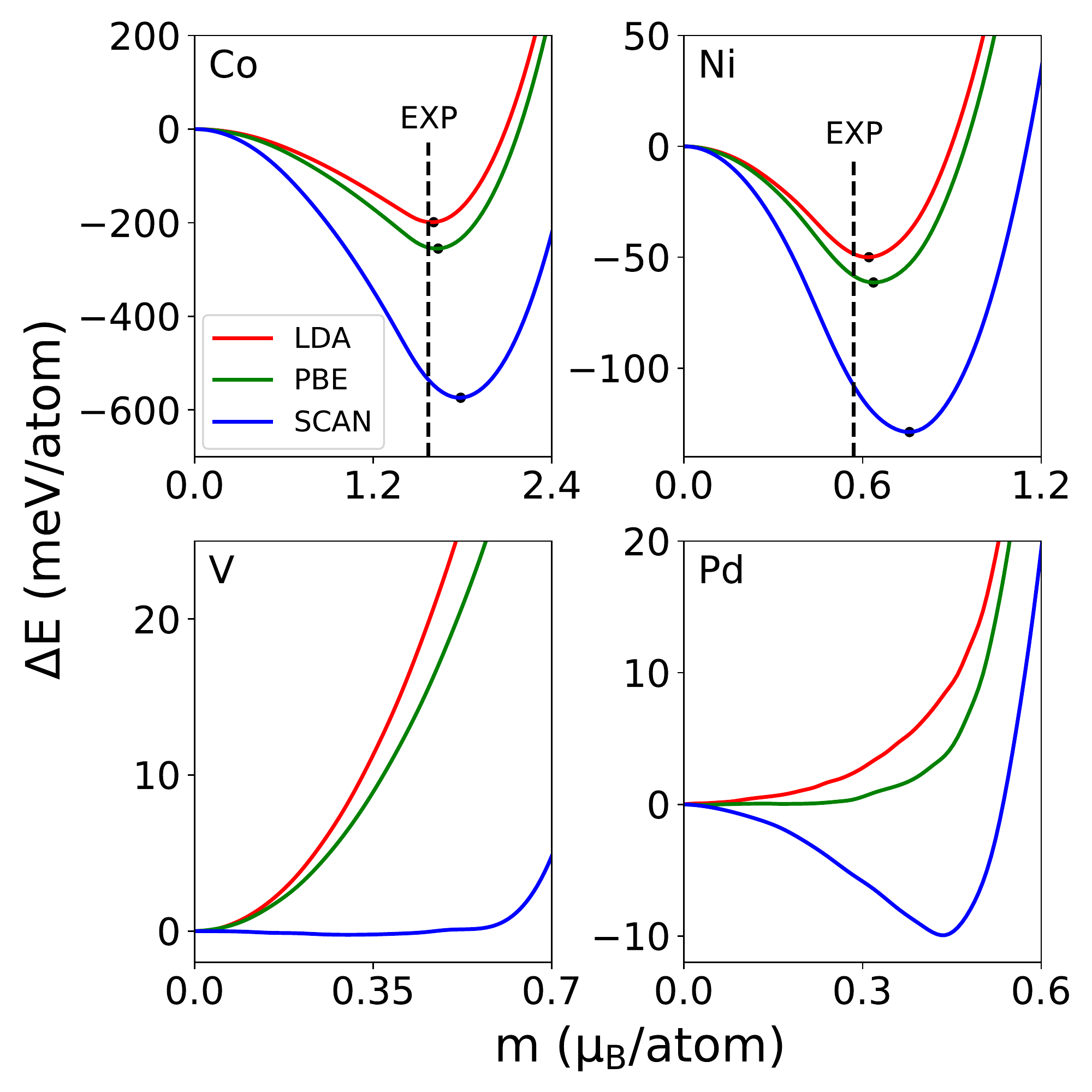}
\caption{FSM calculations of the magnetic energy of transition metal elements
with the LDA, PBE and SCAN functionals.}
\label{other}
\end{figure}

Thus the known data is consistent with good agreement between the
predictions of the PBE functional and experiment.
Importantly, it is inconsistent with the predictions of the SCAN functional.
Specifically, the results point to severe
problems in the SCAN predictions for magnetic energies and moments in Fe.
It is notable that the differences in magnetic energies between SCAN
and the LDA and PBE functionals are much larger than the differences between
predictions of those two functionals.

\begin{table}
\caption{Magnetic data for Ni and Co from fixed spin moment calculations
at the experimental lattice parameters. All quantities are per atom.}
\begin{tabular}{llcccc}
\hline
  &  & ~~~LDA~~~ & ~~~PBE~~~ & ~~SCAN~~ & ~~Expt.~~ \\
\hline
Co & $m_{sp}$ ($\mu_B$) & 1.61 & 1.62 & 1.79 & 1.57 \\
  & $m_{orb}$ ($\mu_B$) & 0.08 & 0.08 & - & 0.15 \\
  & $\Delta$$E_{mag}$ (meV) & 199 & 255 & 574 & - \\
\hline
Ni & $m_{sp}$ ($\mu_B$) & 0.62 & 0.63 & 0.76 & 0.57 \\
  & $m_{orb}$ ($\mu_B$) & 0.05 & 0.05 & - & 0.05 \\
  & $\Delta$$E_{mag}$ (meV) & 50 & 61 & 129 & - \\
\hline
\end{tabular}
\label{tab-other}
\end{table}

Considering the very different predictions of SCAN as compared to standard
functionals for the magnetic properties of Fe, it is of interest to 
investigate whether this is general problem, or if it is restricted to Fe.
Accordingly, we performed fixed spin moment and self-consistent
calculations for other materials.
We start with cementite (Fe$_3$C), which is ferromagnetic and
a key ingredient in many
steels. The calculated spin magnetization per three iron atom
formula unit is 5.75 $\mu_B$ with the PBE functional and 6.87 $\mu_B$
with SCAN (based on self-consistent VASP calculations at the
experimental lattice parameters; very similar values were obtained
from LAPW FSM calculations). This compares with a total room temperature
saturation magnetization of 5.3 $\mu_B$ from experiment,
\cite{hofer}
indicating again a substantial error with SCAN.

Fig. \ref{other}
and Table \ref{tab-other}
give the results of FSM calculations for other
elements with the experimental structures and lattice parameters.
Hexagonal close packed (hcp)
Co and fcc Ni are the other ferromagnetic 3d elements. Ni
is regarded as a prototypical itinerant ferromagnet.
SCAN gives very much larger magnetic energies for these
two elements as compared with PBE and LDA.
We also find enhanced spin moments with SCAN, and similar to Fe we
find significant degradation with respect to experiment for both
Ni and Co. The calculated spin moments with the SCAN functional
are 1.80 $\mu_B$ for Co and 0.77 $\mu_B$ for Ni.
bcc V and fcc Pd are both paramagnetic metals down to 0 K according to 
experiment. Pd is very close to ferromagnetism, and for this reason
exhibits strong spin fluctuations that have been implicated in preventing
a superconducting state in this element.
\cite{berk,larson}
Pd is a particularly interesting test for density functionals, since it
is incorrectly predicted to be ferromagnetic by some hybrid functionals,
\cite{tran-pd}
while showing borderline ferromagnetic behavior with standard GGAs.
\cite{singh-pd,tran-pd}
V is not as close to ferromagnetism
and is a superconductor at low temperature. \cite{roberts}
Our PBE and LDA results are consistent
with these experimental facts. SCAN on the other hand predicts an effectively
infinite susceptibility for V, and a low moment ferromagnetic state
for Pd.
Thus qualitatively similar to Fe, SCAN strongly overestimates the magnetic
tendencies of V, Co, Ni and Pd.

The above results point to a surprising degradation of the predictions of
SCAN relative to PBE in describing magnetism in transition metals, and
suggest caution in the use of this functional for predicting magnetic
properties of materials.
This may perhaps be due to the challenge of obtaining the itinerant
physics of systems like Fe with multiple partially occupied d-orbitals,
and at the same time reproducing correct physics of atoms, including
cancellation of self-interaction.
In any case, we hope
that the above results may motivate further work to
develop improved meta-GGA functionals, particularly functionals
that satisfy known constraints,
from the inhomogeneous electron gas, including the many constraints
satisfied by SCAN, and possibly additional constraints, and at the
same time predict accurate magnetic properties of metals.

This work was supported by
the U.S. Department of Energy, Office of Science,
Basic Energy Sciences, Award Number DE-SC0019114.
We are grateful for helpful discussions with Guangzong Xing.

\bibliography{scan}

\end{document}